# I-BEAT: New ultrasonic method for single bunch measurement of ion energy distribution

Daniel Haffa*[1], Rong Yang[§1], Jianhui Bin[1,4], Sebastian Lehrack[1], Florian-Emanuel Brack[5,6], Hao Ding[2,3], Franz Englbrecht[1], Ying Gao[1], Johannes Gebhard[1], Max Gilljohann[2,3], Johannes Götzfried[2], Jens Hartmann[1], Sebastian Herr[1], Peter Hilz[1], Stephan D. Kraft[5], Christian Kreuzer[1], Florian Kroll[5,6], Florian H. Lindner[1], Josefine Metzkes[5], Tobias M. Ostermayr[1,4], Enrico Ridente[1], Thomas F. Rösch[1], Gregor Schilling[2], Hans-Peter Schlenvoigt[5], Martin Speicher[1], Derya Taray[1], Matthias Würl[1], Karl Zeil[5], Ulrich Schramm[5,6], Stefan Karsch[2,3], Katia Parodi[1], Paul R. Bolton[1], Walter Assmann[1] and Jörg Schreiber[1,3]

[1]*Lehrstuhl für Medizinphysik, Fakultät für Physik, Ludwig-Maximilians-Universität München, 85748 Garching b. München, Germany*

[2]*Lehrstuhl für Experimentalphysik - Laserphysik, Fakultät für Physik, Ludwig-Maximilians-Universität München, 85748 Garching b. München, Germany*

[3]*Max-Planck-Institut für Quantenoptik, 85748 Garching b. München, Germany*

[4]*Accelerator Technology and Applied Physics Division, Lawrence Berkeley National Laboratory, Berkeley, CA 94720, USA*

[5]*Helmholtz-Zentrum Dresden–Rossendorf (HZDR), Bautzner Landstr. 400, 01328 Dresden, Germany*

[6]*Technische Universität Dresden, 01062 Dresden, Germany*

Email: *Daniel.Haffa@physik.lmu.de, [§]Rong.Yang@physik.lmu.de

**The shape of a wave carries all information about the spatial and temporal structure of its source, given that the medium and its properties are known. Most modern imaging methods seek to utilize this nature of waves originating from Huygens' principle[1,2]. We discuss the retrieval of the complete kinetic energy distribution from the acoustic trace that is recorded when a short ion bunch deposits its energy in water. This novel method, which we refer to as Ion-Bunch Energy Acoustic Tracing (I-BEAT), is a generalization of the ionoacoustic approach[3–5]. Featuring compactness, simple operation, indestructibility and high dynamic ranges in energy and intensity, I-BEAT is a promising approach to meet the needs of petawatt-class laser-based ion accelerators[6–9]. With its capability of completely monitoring a single, focused proton bunch with prompt readout it, is expected to have particular impact for experiments and applications using ultrashort ion bunches in high flux regimes. We demonstrate its functionality using it with two laser-driven ion sources for quantitative determination of the kinetic energy distribution of single, focused proton bunches.**

Laser-plasma accelerator development has been advancing rapidly in the past few decades, opening a new frontier in accelerator physics. High particle numbers at a broad range of relativistic energies, originating from an exceptionally confined region in space and time, are some of the outstanding features of laser-plasma based ion accelerators[6–9]. Tremendous efforts and progress regarding increasing intensity, repetition rate and various target refinements bring many applications within reach of today's capability. This impressive progress enhances the need for innovative diagnostics development. The direct measurement of the ion kinetic energy distributions that satisfy emerging online evaluation requirements, such as high repetition rate detection at increased ion energies while being robust and EMP (electromagnetic pulse) resistant, is the primary motivation for this work.

Volumetric detectors such as stacks of radiation-sensitive films[10] and scintillators[11,12] allow recording of the energy-dependent angular distribution across the full ion bunch energy spectrum. Other methods to date typically rely on sampling a minor fraction of an ion bunch in magnetic and Thomson parabola spectrometers[13–15]. Also, renewed efforts for collecting a large portion of the diverging ion bunch[16–20] have proven successful. Due to the energy selectivity typical of particle optics, the need arises to reliably characterize the particles´ full energy distribution from a single bunch at application sites[21] ideally with direct, prompt single bunch readout.

Here we report on a new method for ion energy measurements, relying on analysis of the acoustic wave, generated when an ion bunch dissipates its energy into a water volume[3,5,22]. The short bunch duration and therefore intense particle flux of laser-accelerated protons allows for the first time a reconstruction of the depth dose distribution and therefore the complete energy distribution of a single proton bunch (without any averaging or scanning as required in previous work[22]).

I-BEAT consists of two parts, the detector that measures the acoustic traces (Fig. 1, details can be found in supplementary material) and the reconstruction algorithm that yields the complete energy distribution. As seen in Fig. 1a the proton bunch enters the cylindrical water chamber from the left (4 cm diameter and length of 10 cm), through an 11 µm thick titanium foil with 1 cm diameter. The ultrasound transducer (Videoscan V311, Olympus), operating in the MHz regime, records the acoustic waves propagating towards the on-axis transducer. If the bunch duration is much shorter than the typical duration of the acoustic wave period (i.e. on the order of µs) the energy deposition can be considered instantaneous. The pressure signal on the axis of propagation at a distance $z = z_d$ (position of the ultrasound detector) is then obtained by solving the wave equation (see supplementary material) as follows

$$p(z_d, t) = \frac{\Gamma n_i}{4\pi c \sigma_r^2} \frac{\partial}{\partial t} \int_{z_d - ct}^{z_d + ct} B_s(z')\, e^{-\frac{1}{2\sigma_r^2}\left[c^2 t^2 - (z_d - z')^2\right]} dz', \tag{1}$$

where $B_s(z') = \int B(E_{kin}, z') f(E_{kin})\, dE_{kin}$ represents the instantaneously generated spread out Bragg curve, $f(E_{kin})$ corresponds to the normalized kinetic energy distribution of $n_i$ ions in a single bunch, $c$ is the speed of sound in the medium in which the ions are stopped and $\Gamma$ is the Grüneisen parameter[2]. For simplicity we assume the transverse profile of the ion bunch in water to be Gaussian with standard deviation $\sigma_r$.

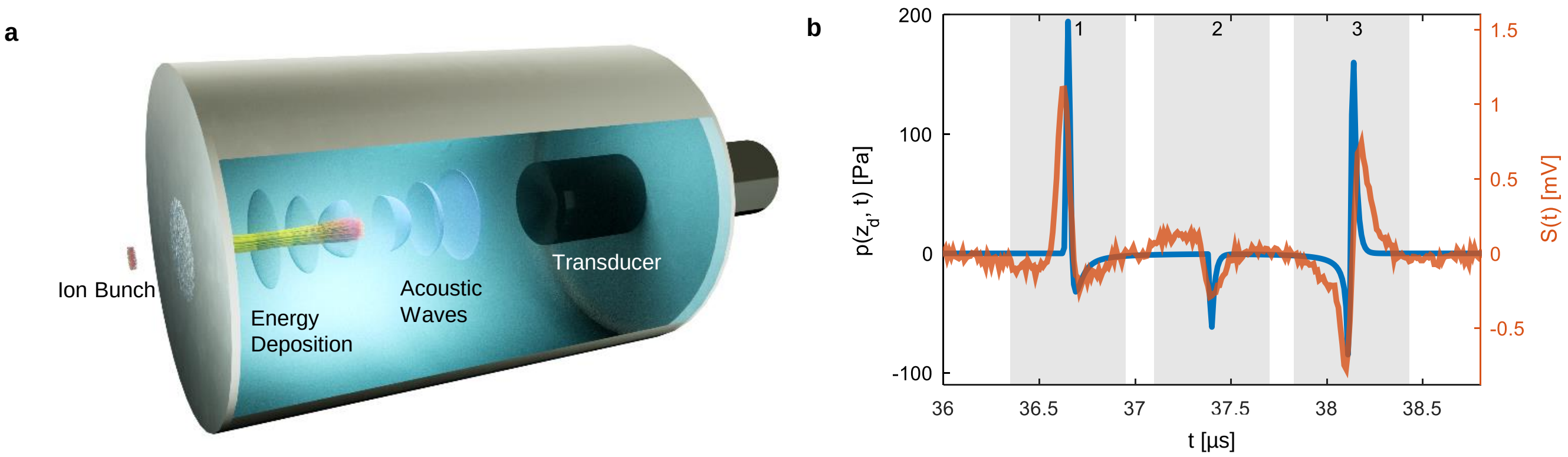

**Figure 1 | Experimental scheme of I-BEAT. a,** The short ion bunch enters the water volume via an 11 µm thick titanium foil (1cm diameter), depositing its energy in the water and generating an acoustic wave, which is measured via a transducer. This generates a signal as shown in **b**, where the orange curve is an example trace for a mono-energetic 9.4 MeV proton bunch measured at the MLL Tandem accelerator and the blue curve represents simulated results, considering an ideal detector with equal conditions.

The blue curve in Fig. 1b exemplarily shows the pressure wave that would be recorded from a mono-energetic 9.4 MeV proton bunch. The signal can be viewed in three segments. In addition to the pulse that propagates directly to the transducer and arrives 36.5 µs after proton impact (segment 1), a second pulse initially propagates in the opposite direction towards the entrance window, and is subsequently reflected back towards the transducer (segment 3). The combination of direct and reflected acoustic signals (segments 1 & 3) images the pressure source from two sides and can thus even be interpreted as a first step towards tomographic analysis. The smaller intermediate signal (segment 2) originates from the energy deposition at the entrance window. Reflections from the side walls of the water housing are temporally well separated and highly attenuated and do not disturb the signal.

Direct comparison and mismatch between the measured (orange) and the ideal pressure signal predicted by equation (1) (blue) reveal the need for applying the detector response[23] correction, i.e. the detector transfer function. Therefore a calibration was performed at the MLL Tandem accelerator at Garching, using well defined proton bunches of 40 ns duration with an energy of 9.4 MeV ($dE/E = 10^{-4}$) (see supplementary). The calibration allows the determination of the expected observed acoustic trace $S_m$(t) in volts for a given energy distribution $f(E_{kin})$. The I-BEAT reconstruction algorithm varies $f(E_{kin})$, calculates $S_m$(t) and compares it to the measured curve. With this so called simulated annealing[24] (see methods), we can iteratively retrieve the discretized energy distribution function $f(E_{kin})$ and the transverse bunch size $\sigma_r$ (see supplementary). A first successful I-BEAT reconstruction of the energy distribution of proton beams with a narrow energy spread has also been shown at the tandem accelerator (see supplementary).

After calibration and characterization, I-BEAT was demonstrated using laser-accelerated ion bunches at the Laboratory of Extreme Photonics (LEX Photonics) in Garching near Munich (Fig. 2a)[25]. The ATLAS300 is a Ti:Sa laser system delivering 2.2 J energy (on target) within 30 fs at a central wavelength of 800 nm and repetition rate of 1 Hz with $10^{20}\,\frac{W}{cm^2}$ (on target). By focusing it onto a 250 nm gold foil, a proton bunch with a typically broad TNSA spectrum up to 9 MeV emerged from the surface contamination layers of the plasma source[6,7,26]. A permanent magnetic quadrupole (PMQ) doublet[27,28] placed closely behind the proton source collected a large portion of this bunch and focused it to the application site outside of the vacuum chamber[21]. The PMQ doublet chromaticity was exploited to focus the design energy to a desired position (application site) by adjusting both the distances to the target and each component of the doublet. The energy distribution of the proton bunch at the focal position was thereby filtered (i.e. narrowed down to a range around the design energy as depicted in Fig. 2d) and not measurable with existing high repetition rate techniques. I-BEAT, placed at application site, enables this measurement. The bunch exited the vacuum chamber through a 50 µm Kapton window (1 cm wide in horizontal and 5 cm in vertical dimensions), traversed 3.3 cm of air and the 11 µm thick titanium detector entrance window before reaching the water volume. An additional dipole magnet with an effective field of 150 mT over a length of 0.1 m was employed to 'clean' the signal, i.e. directing potential contamination attributed to energetic electrons and low energy ions away from the detector entrance[14].

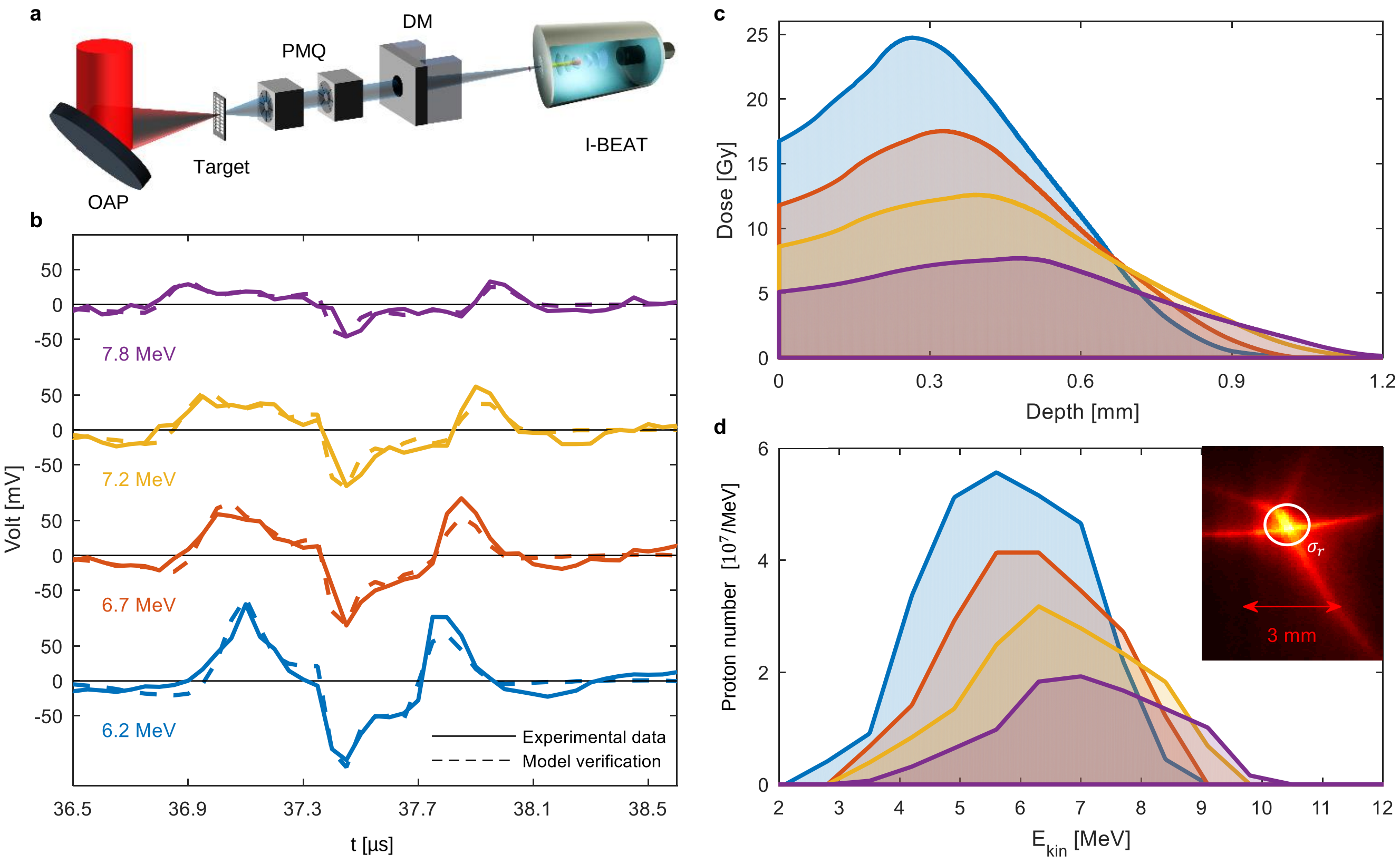

**Figure 2 | Results for laser-accelerated proton bunches: a**, Shows the schematic of the experiment setup. A high power laser (ATLAS300) is focused with an off-axis parabola (OAP) onto a foil target. Two permanent magnet quadrupoles (PMQ) are used to focus a short ion bunch. A dipole magnet (DM) is used to remove electrons and low energy ions from the swift ion bunch, which is focused within the ionoacoustic detector. **b,** Acoustic signals of single proton bunches. The design energies attenuated to 6.2, 6.7, 7.2 and 7.8 MeV on entering the water volume, are set by positioning of the PMQs. The solid line is the measured acoustic signal and the dashed line corresponds to the calculated signal from the retrieved spectrum in **d**. **c**, Depth dose curves corresponding to the different energy settings. The dose on the central axis is $D(z) = \frac{1}{\rho} \cdot \frac{1}{2\pi\sigma_r^2} \cdot B_s(z)$. **d**, Absolute proton energy distributions of single proton bunches of the different design energy settings in the ion focus. The inset reveals a focal plane image of a single proton bunch at the position of the detector entrance, taken with an image plate. The $\sigma_r$ is the resulting Gaussian width of I-BEAT.

Acoustic traces corresponding to single proton bunches with design energies attenuated to 6.2, 6.7, 7.3 and 7.8 MeV at the water tank are presented in Fig. 2b. The spot size at the focus of the laser-accelerated proton bunch was not well defined (a picture taken with an image plate can be seen in the inset of Fig. 2d). While the highest dose is located in an area smaller than $\sigma_r = 1.5 \pm 0.2$ mm (error is due to the step size of the fitting algorithm) it features also radial caustic shapes at lower dose[29]. Although this overall shape is not Gaussian, the evaluation of the residuals $\Sigma$ (see methods) for different $\sigma_r$ produced a comparable result as long as $\sigma_r$ less than 2.5 mm was chosen. To accelerate the reconstruction process $\sigma_r$ was fixed to 1.5 mm for the reconstruction of the different energy settings, since this was the best fit for reconstruction of the 6.2 MeV case (see supplementary material).

Retrieved absolutely calibrated proton energy distributions in the PMQ doublet focal plane are presented in Fig. 2d. The proton number of $10^7$ may seem small but considering that the bunch length is ns and the Gaussian width is 1.5 mm the proton flux is intense (bunch current of µA, assuming a ns bunch duration). As part of the numerical reconstruction, those final retrieved energy distributions ($n_i f(E_{kin})$ ) are used to calculate an expected signal $S_m$(t), employing equation (1) and the transfer function (dashed curves in Fig. 2b). The excellent conformity of the final retrieved and measured signals demonstrates successful reconstruction of the ion energy distribution. The corresponding on-axis depth dose distribution in Fig. 2c is of particular interest for biomedical application and is a natural byproduct of I-BEAT.

We note that there currently exists no other established method, to which we could reliably compare our results at the presented proton energies (~7 MeV) with similar energy resolution in a focused beam. Therefore, we conducted another experiment at the petawatt laser acceleration facility in Dresden (Draco)[30], enabling a direct comparison of I-BEAT to the well-established radiochromic film (RCF) stack detector and further allowing a demonstration of the feasibility of I-BEAT at a higher fluence level approaching $10^9$ per bunch. Particle numbers beyond $10^{10}$ per bunch are foreseen as realistic (see supplementary).

The Draco laser is a Ti:Sa petawatt laser system, capable of delivering an energy of up to 30 J on target within 30 fs at a repetition rate of 1 Hz. Here it was operated at reduced, 12 J on target, due to a temporal pulse cleaning via a plasma mirror[31]. It was focused onto a 200 nm thin plastic foil, generating a proton bunch with a typically broad TNSA spectrum with energies up to 30 MeV. A pulsed solenoid[20,32] was used to focus a design energy into the detector outside of the vacuum chamber[33] (Fig. 3a, see methods). This time, energies up to 30 MeV in this beam time enabled a comparison with an RCF stack (EBT3 Gafchromic film, calibrated with an X-ray tube). Fig. 3 shows a shot for a design energy of about 16 MeV and its comparison to an RCF stack. This energy setting has been chosen to optimize the signal-to-noise ratio recorded by the I-BEAT detector. Fig. 3b shows the measured signal and the calculated signal using the evaluated spectrum of Fig. 3c as an input. Fig. 3d validates that I-BEAT can reconstruct the depth dose distribution quantitatively. The depth resolution (horizontal spacing between data points) of I-BEAT is due to the sampling rate and transfer function and the error bar due to the limited band width of the transducer (10 MHz). The error bars of the dose result from fluctuation of the particle number per bunch of the Tandem accelerator during the calibration. Fitting a 2D Gaussian to the Gafchromic film with the highest dose yields $\sigma_{\mathrm{h}}$ of 3.6 mm and $\sigma_{\mathrm{v}}$ of 2.2 mm with an average of 2.8 mm. The fitting result of I-BEAT with $\sigma_{\mathrm{r}} = 3.0 \pm 0.2$ mm matches and shows that the transversal information can be retrieved with a single transducer. The energy resolution of I-BEAT at this stage is already better than that of the RCF stack (spatial resolution is nearly doubled, Fig. 3d). In general, the longitudinal spatial resolution of I-BEAT is limited by the frequency of the detector (transducer) and the transfer function. Since the spatial resolution, to first order, is constant along the propagation direction (Fig. 3d), the energy resolution of I-BEAT intrinsically increases for higher ion energies (Fig. 3c) and thus outmatches techniques based on magnetic deflection[13–15], where the energy resolution decreases at higher ion energies. The energy resolution approaching high kinetic energies is likely limited by energy loss straggling[34] as for all depth range monitors such as RCF stacks (in general termed position sensitive calorimeters).

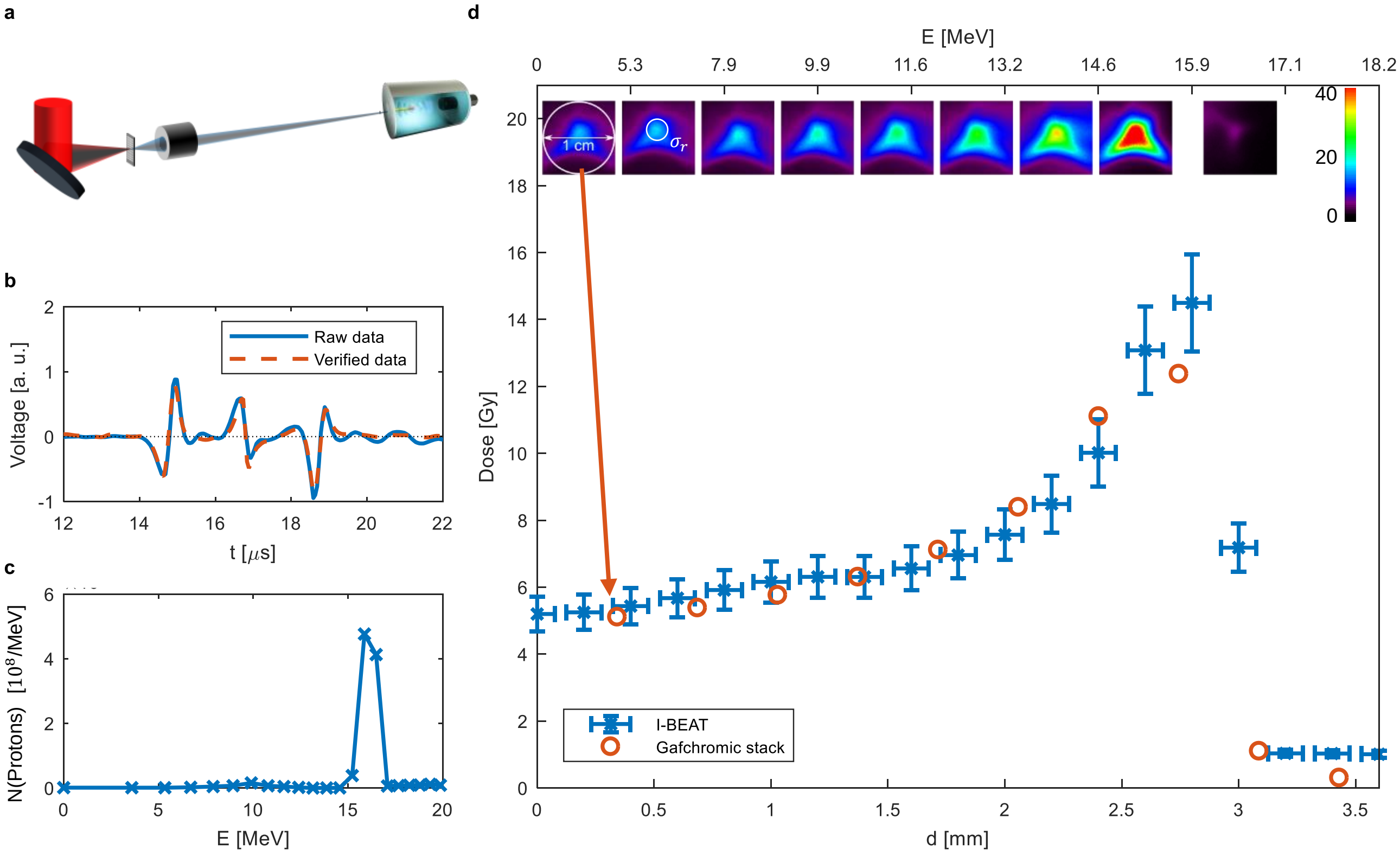

**Figure 3 | Comparison to RCF stacks at Draco (Dresden): a,** is a sketch of the setup. A pulsed solenoid was used to focus a certain design energy. **b,** is the measured signal and the calculated signal according to the energy distribution evaluated in c. **c,** shows the reconstructed proton spectrum. **d,** is the depth dose distribution determined by I-BEAT compared to the one obtained by an RCF stack. The corresponding layers of the stack are depicted and the dose (colour coded in Gy) is evaluated over a circle of 1 cm diameter (entrance of the detector) for both the RCF stack and I-BEAT. The evaluated Gaussian profile determined by I-BEAT yields $\sigma_r = 3.0 \pm 0.2$ mm (shown in the second film picture). The upper axis shows the corresponding proton energy of different penetration depths. The error bar on the y-axis is due to fluctuation of the particle number per bunch of the Tandem accelerator during the calibration.

We have demonstrated the potential of I-BEAT as a novel, compact and simple method for characterizing the absolute kinetic energy distribution of single ultrashort proton bunches. Even though we used the detector in air, an operation in vacuum is possible without modification and has already been tested (see supplementary material). As in most depth dose monitors, the energy range in which I-BEAT operates can be easily adapted by proper choice of medium and size to accommodate the complete Bragg curve (starting from several MeV as presented in Fig. 2 up to several 100 of MeV protons[34]). The harsh conditions typically encountered near laser-plasmas (notably strong electromagnetic pulses[35]) are typical key challenges in the design and evaluation of online electronic detectors. In contrast, the relatively low speed of sound results in an inherent µs delay of the acoustic signal, which effectively stores the information. This allows ample time for the decay of prompt undesirable artifacts of the intense laser-plasma interaction, rendering I-BEAT measurements unaffected. We demonstrated the reconstruction of the complete energy distribution of a single proton bunch using one single transducer (no averaging, no scanning as required in previous work[22]). Our pioneering demonstration with a single transducer uniquely enables single bunch reconstruction of an otherwise unknown energy distribution.

While short, intense ion bunches typically saturate detectors (or even cause radiation damage), I-BEAT, using water as a medium, is nearly indestructible and offers a high dynamic range from $10^7$ (Fig. 2) up to $10^{11}$ (see supplementary material) protons/$mm^2$ for intense ion bunches. Further key advantages of I-BEAT include compactness, robustness, simplicity of operation and low cost. I-BEAT is a high-repetition-rate system with a good energy resolution also at high energies[36] and offers the possibility of dose control at application sites, especially for guided and focused ion beams. In conclusion, those key features have been found to complement or outperform established techniques. Thus I-BEAT seems a promising approach satisfying the needs of the existing and upcoming high repetition rate petawatt laser facilities (ELI, GIST[37], Apollon[38], CALA[39], Draco[30]), in particular for monitoring bunches downstream of transport and focusing optics. The duration of the I-BEAT signal remains smaller than twice the range of ions divided by the speed of sound, for 100 MeV protons this is 100 µs. The I-BEAT detector could thus be advanced to an operation with repetition rates up to kHz.

By employing more than one transducer in additional dimensions the extension to a 3D-tomographic configuration seems feasible. In combination with magnetic energy selection optics (including quadrupoles or solenoids) the detailed diagnosis of more complicated particle bunches consisting of mixed species (for example protons and carbons), can be incorporated in the reconstruction algorithm (see supplementary). In the field of biomedical ion irradiation, measuring the dose is crucial. While most commonly used detection methods for ions rely on energy deposition in the detector, I-BEAT in principle will allow direct measurement of the dose, even when the energy is completely deposited within a biomedical sample or used for another application[40], by measuring the generated sound wave, as with ultrasound imaging. Thus I-BEAT, as a versatile spectrometer, could be advanced to an in-vivo dosimeter[41,42] and enhance imaging applications that use energetic short ion bunches.

# Methods

### Simulated Annealing

The method of simulated annealing relies on a random variation of an initial spectrum $f_i(E_{kin})$ to obtain a modified spectrum $f_m(E_{kin})$, and comparing the two predicted acoustic signals $S_i(t), S_m(t)$ from the initial and the modified inputs with the measured signal $S_0(t)$ by the least squares method. If the residual $\Sigma_m$ is smaller than $\Sigma_i$, the algorithm continues with the modified spectrum as the updated input distribution for the next cycle. Otherwise, with probability $exp(-(\Sigma_m - \Sigma_i)/T)$, the modified spectrum is rejected, while with the probability $1 - exp(-(\Sigma_m - \Sigma_i)/T)$ the intial spectrum is taken into the next cycle to avoid being caught in a local minimum. $T$ is the annealing schedule temperature and is set to 1. After a sufficient amount (few hundred) of iterations, $\Sigma_i$ converges to a global minimum value, the final residual.

### Experimental Setup at the Draco Laser

In the experiment, Draco delivered 30 fs pulses with an energy of about 12 J with enhanced temporal contrast using a re-collimating single plasma mirror on target. Using plastic foil targets with a thickness of about 200 nm, the laser drives a TNSA proton source with cut-off energies in the range of 30 MeV. The tunable solenoid magnet[20,32] is positioned 80 mm behind the target and is therefore able to collect the high energetic part of the beam without particle loss. It acts as chromatic lens and can be used to generate a focus of a desired mean energy at the irradiation site in air about two meters downstream of the target. The energy bandwidth of the transported proton bunch amounts to about 20% (FWHM) at the focus position. For the presented experiment, a mean proton energy of 15.4 MeV was focused into the I-BEAT detector, corresponding to a solenoid current of 12 kA leading to a magnetic field of ca. 10 T, accordingly. This proton energy has been chosen to optimize the signal-to-noise ratio recorded by the I-BEAT detector.

**Data availability**

The data that support the plots within this paper and other findings of this study are available from the corresponding authors upon reasonable request.

## Acknowledgements

The DFG-funded Cluster of Excellence Munich-Centre for Advanced Photonics (MAP) supported this work. We thank the support of the operation team of the MLL Tandem accelerator and the ATLAS300 laser system. We thank S. Zherlytsin, T. Herrmannsdoerfer for useful discussions and the development of the pulsed solenoids and the Draco

Laserteam S. Bock, R. Gebhardt, U. Helbig, T. Pueschel for excellent laser support. F.H.L. was supported by the BMBF under contract 05P15WMEN9. T.F.R. is scholar of the German Academic Scholarship Foundation. J. H. B. is funded by a Feodor Lynen Fellowship of the Alexander von Humboldt Foundation. M. W. and H. D. acknowledge financial support from the IMPRS-APS. J. Ge. acknowledges financial support from the Hanns Seidel Foundation. Partially supported by EC Horizon 2020 LASERLAB-EUROPE/LEPP (Contract No. 654148).

## Author Contribution

D. H., C. K., T. M. O., and S. L. built the experimental infrastructure in LEX Photonics and designed the laser plasma experiment. D. H., R. Y., J. H. B., M. S., J. H., S. L., T. M. O., C. K. , F. H. L., M. W, E. R., T. F. R., F. E., J. Ge. and Y. G. performed the experiment at LEX Photonics. T. F. R., J. H., S. H., P. H., and F. H. L. designed and prepared the magnetic transport system for the proton beam. S. L., R. Y., D. H., and W. A. designed and developed the I-BEAT detector. H. D., J. Gö., M. G., G.S., and S. K. built and operated the ATLAS300 Laser. D. H., R. Y., and S. L. performed the experiment at the MLL Tandem accelerator. U. S., K. Z., J. M., S. D. K., F. K., F. E. B. and H. P. S. developed and built the Draco laser-ion-acceleration system including target, plasma mirror and pulsed solenoids. K. Z., F. K., F. E. B., H. P. S., D. H., R. Y., S. L. performed the experiment at Draco. R. Y., J. S., D. H., and J. H. B. analyzed the data, discussed and interpreted the results and prepared the manuscript. J. S., K. P., W. A. and P. R. B. conceived and supported the overall project and provided final input to the manuscript.

## Competing financial interest

The authors declare no competing financial interests.

## Additional information

Correspondence and requests for materials should be addressed to D. H. or R. Y.

# Supplementary Material for

## I-BEAT: New ultrasonic method for single bunch measurement of ion energy distribution

Daniel Haffa*[1], Rong Yang[§1], Jianhui Bin[1,4], Sebastian Lehrack[1], Florian-Emanuel Brack[5,6], Hao Ding[2,3], Franz Englbrecht[1], Ying Gao[1], Johannes Gebhard[1], Max Gilljohann[2,3], Johannes Götzfried[2], Jens Hartmann[1], Sebastian Herr[1], Peter Hilz[1], Stephan D. Kraft[5], Christian Kreuzer[1], Florian Kroll[5,6], Florian H. Lindner[1], Josefine Metzkes[5], Tobias M. Ostermayr[1,4], Enrico Ridente[1], Thomas F. Rösch[1], Gregor Schilling[2], Hans-Peter Schlenvoigt[5], Martin Speicher[1], Derya Taray[1], Matthias Würl[1], Karl Zeil[5], Ulrich Schramm[5,6], Stefan Karsch[2,3], Katia Parodi[1], Paul R. Bolton[1], Walter Assmann[1] and Jörg Schreiber[1,3]

[1]*Lehrstuhl für Medizinphysik, Fakultät für Physik, Ludwig-Maximilians-Universität München, 85748 Garching b. München, Germany,*

[2]*Lehrstuhl für Experimentalphysik - Laserphysik, Fakultät für Physik, Ludwig-Maximilians-Universität München, 85748 Garching b. München, Germany,*

[3]*Max-Planck-Institut für Quantenoptik, 85748 Garching b. München, Germany.*

[4]*Accelerator Technology and Applied Physics Division, Lawrence Berkeley National Laboratory, Berkeley,CA 94720, USA.*

[5]*Helmholtz-Zentrum Dresden–Rossendorf (HZDR), Bautzner Landstr. 400, 01328 Dresden, Germany*

[6]*Technische Universität Dresden, 01062 Dresden, Germany*

This supplementary materials addresses details of I-BEAT that have not been discussed in the paper. The detector itself is described and pictures are shown. The calculation leading to equation (1) is outlined. The calibration and first experiments of I-BEAT, performed at the Tandem accelerator in Garching near Munich[1], is described and the results are shown and discussed. The dynamic range of I-BEAT and thus its behaviour at higher particle numbers is further estimated. The data analysis for the retrieval of the energy spectrum is investigated, especially the influence of different $\sigma_r$ in the situation of the experiment at LEX Photonics is described. Simulations, where I-BEAT is implemented in typical circumstances are shown.

## Detector Setup

An ionoacoustic detector relies on the detection of the acoustic signal that is generated due to thermal heating of an ion bunch dissipating its kinetic energy in water. Since operation in vacuum was desired, we chose a KF40 vacuum pipe with 10 cm length as water container. A hole of 1 cm diameter at the front plate is covered with an 11 µm thick titanium foil that is airtight and waterproof and functions as an entrance window for the ion bunch. The transducer was attached to the rear flange and positioned in the water sample. We chose a focusing transducer (focal length of 25.4 mm) to enhance the signal. Signals generated in focal distance will have the best temporal resolution while the resolution drops off out of focus. Supplementary Fig. 1c shows the geometry of the detector used at the Laberoratory of Extreme Photonics (LEX Photonics). A picture of the transducer is given in Supplementary Fig. 1d and the used amplifier (60 dB, HVA-10M-60-B, FEMTO Messtechnik GmbH) in Supplementary Fig. 1e. Note that all parts in the electronic chain influence the signal response and have to be included in the calibration. Since the motivation for I-BEAT was its implementation in a laser-plasma ion accelerator a setup-picture is shown in Supplementary Fig. 1f. The picture is taken at LEX Photonics in Garching near Munich. It shows the implementation of I-BEAT inside the vacuum chamber. The detector was modified at the experiment at the Draco laser. The length of the tube and thus the distance of the source of the generated sound signal was shortened and now positioned directly in the focal plane of the transducer. This improves the signal-to-noise ratio and the temporal resolution of the detector. Supplementary Fig. 1h is a picture of the setup at the Draco laser.

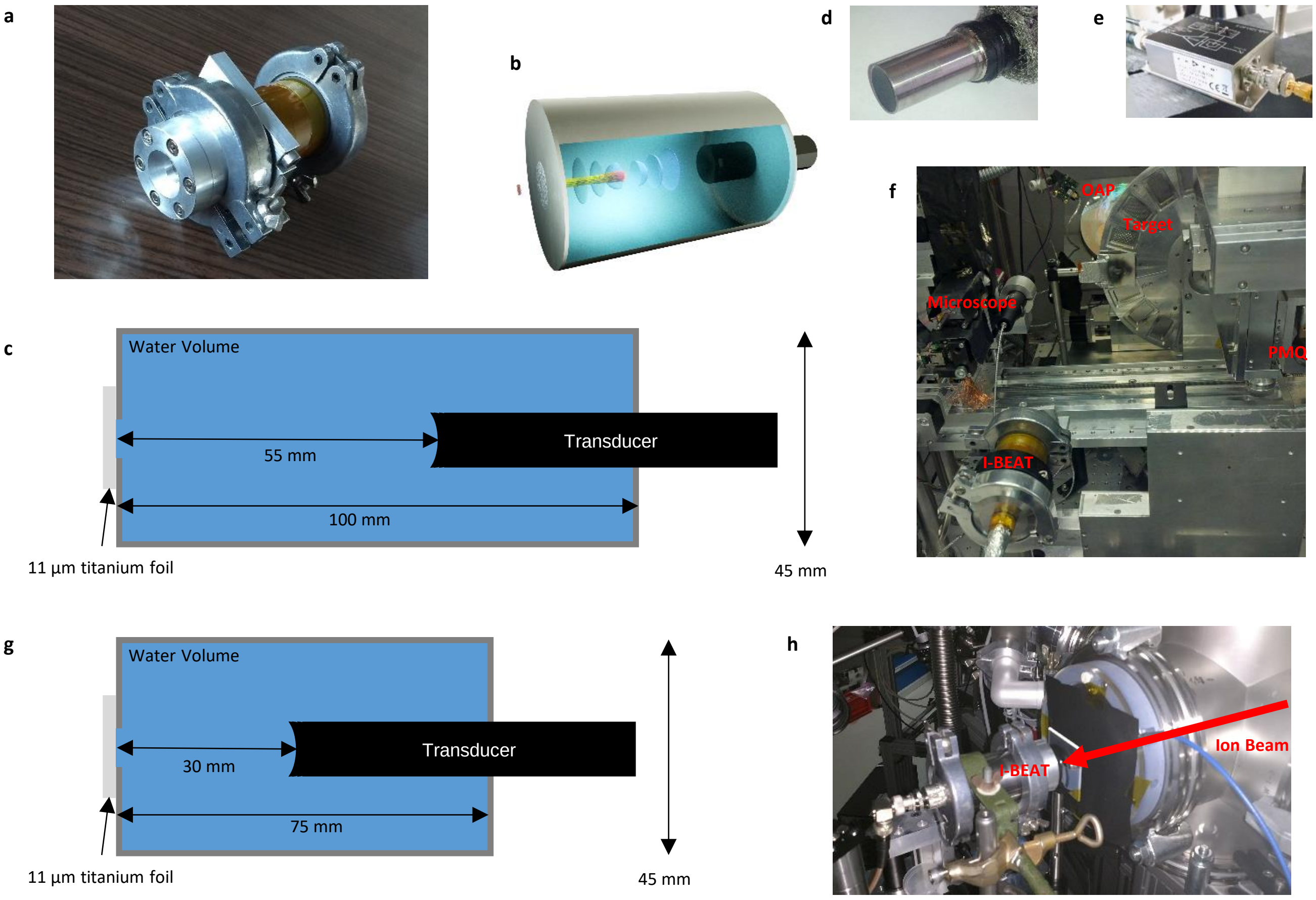

**Supplementary figure 2 | Setup of the ionoacoustic detector. a,** is a picture of the tube accommodating the water sample (frontview). **b,** shows the *detector and depicts the physical process of I-BEAT.* ***c*** *is a sketch emphasizing the dimensions of the used detector at LEX-Photonics.* ***d,*** *is a picture of the transducer that was used during the experiments. It is a focusing transducer with a mean frequency of 10 MHz.* ***e,*** *is the Voltage amplifier used in the experiment.* ***f,*** *shows the setup at the laboratory of extreme photonics in Garching. This shows the implementation of I-BEAT directly in the vacuum chamber. The laser was focused with the off-axis parabola (OAP) onto the target. Two permanent magnet quadrupoles were used to focus the proton bunch into the water volume.* ***g,*** *shows the modified detector that was used at the Draco laser. The detector was shortened in order to accommodate the Bragg peak in the focus of the transducer.* ***h,*** *is a picture of the setup at the Draco laser.*

## Calculation

This part describes a more detailed derivation of equation (1)[2,3]. Also the derivation of the reflection coefficient is explained in detail.

By solving the wave equation

$$\left(\nabla^2 - \frac{1}{c^2}\frac{\partial^2}{\partial t^2}\right)p(\vec{r},t) = \frac{-\Gamma}{c^2}\frac{\partial}{\partial t}H(\vec{r},t) \tag{1}$$

with the Grüneisen-parameter $\Gamma$ in Pa/(J/m³) and the phase velocity of the acoustic wave $c$, we can approximate the heating function $H(\vec{r}',t') = H_s(\vec{r}')\delta(t')$. This separation is valid since the ion energy deposition can be considered as instantaneous. Denoting the Bragg curve produced by a single ion with a specific initial kinetic energy $E_{kin}$ with $B(E_{kin},z')$ (in *J/m*) and considering a transverse Gaussian distribution with cylindrical symmetry and standard deviation $\sigma_r$, the solution on axis at the detector position $z = z_d$ becomes

$$p(z_d,t) = \frac{\Gamma n_i}{4\pi c\sigma_r^2}\frac{\partial}{\partial t}\int_{z_d-ct}^{z_d+ct} B_s(z')\, e^{-\frac{1}{2\sigma_r^2}\left[c^2t^2-(z_d-z')^2\right]}dz', \tag{2}$$

where $B_s(z') = \int B(E_{kin},z')f(E_{kin})\,dE_{kin}$ represents the instantaneously generated spread out Bragg curve and $f(E_{kin})$ corresponds to the normalized kinetic energy distribution of the number of ($n_i$) ions in a single bunch. The reflectivity of the sound wave at the entrance foil is defined by $R = (Z_M - Z_W)/(Z_M + Z_W)$ with

$$Z_M = Z_0\frac{Z_{load} - iZ_0 tan(k_0d_0)}{Z_0 - iZ_{load}tan(k_0d_0)}, \tag{3}$$

where $k_0 = 2\pi f/c_0$ with $c_0$ and $Z_0 = \rho_0 c_0$ the speed of sound and sound impedance of the mirror material, and $\rho_0$ its mass density. $Z_{load} \approx 0$ (vacuum or air), so that $Z_M \approx -iZ_0 tan(k_0d_0)$ and $R = -e^{i\phi}$ with

$$\phi = atan\left(\frac{2Z_wZ_0tan(k_0d_0)}{Z_W^2 - Z_0^2tan^2(k_0d_0)}\right). \tag{4}$$

In the setting relevant for our case $f < 10\ MHz$ and $d_0 = 11\mu m$, $k_0d_0 < 0.13$. The phase term of the reflectivity can thus be neglected and the reflection $R = -1$ corresponds to that of a fixed end such that the polarity of the reflected wave packet is inverted.

## Calibration

I-BEAT relies on measuring the acoustic traces originating from pressure changes induced by ions in water. The so called transfer function $T(f)$ connects the pressure waves $p(t)$ with the measured acoustic signals $S_m(t)$ by

$$T(f) = \frac{FT[S_m(t)]_f}{FT[p(t)]_f} \qquad (5)$$

and fully depends on the employed transducer and configuration. Thus the transfer function for our detector in units *V/Pa* for our detector had to be calibrated first (Supplementary Fig. 2b)[4].

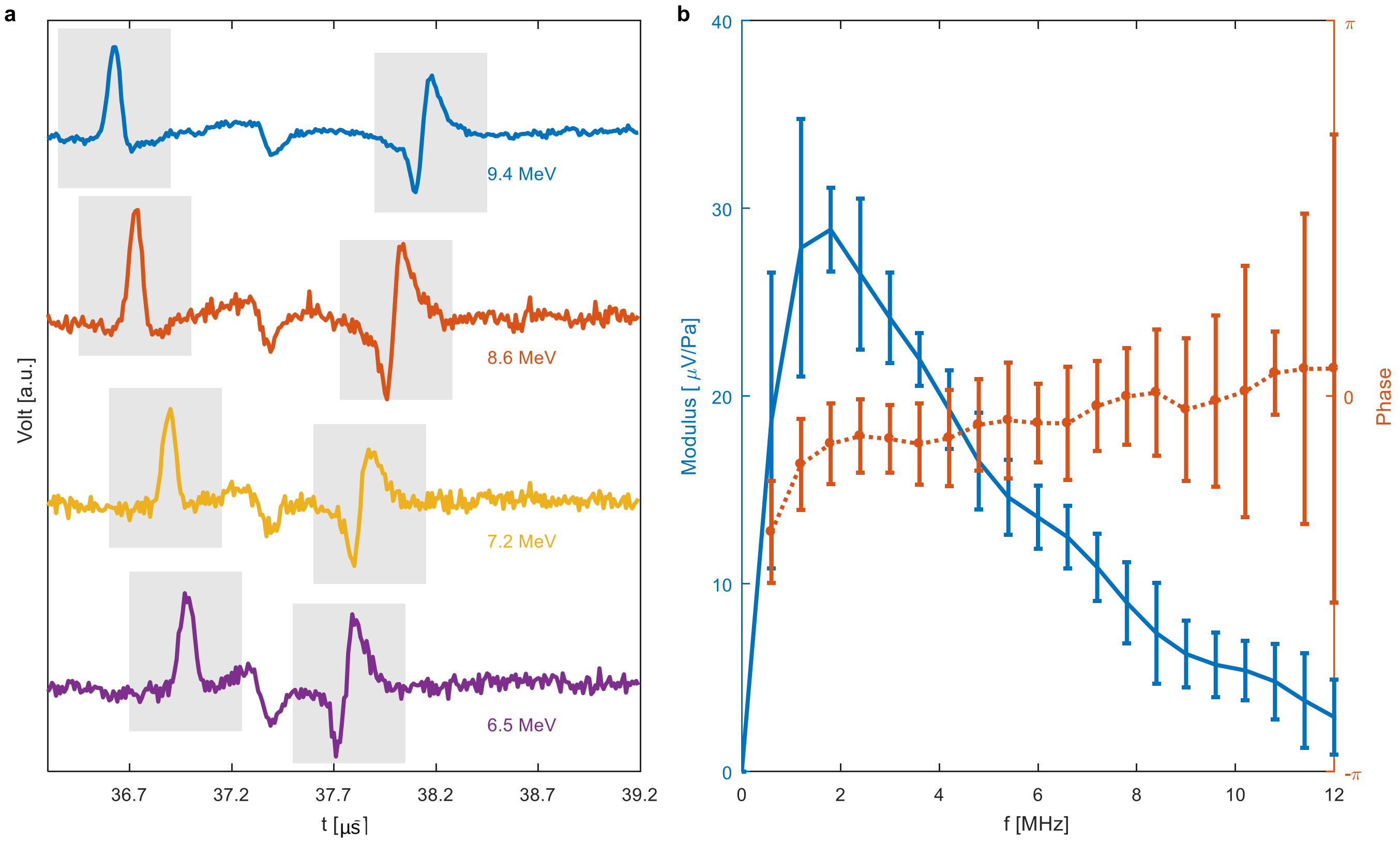

**Supplementary figure 2 | Evaluation of the transfer function. a,** shows acoustic traces recorded with proton bunches at the Tandem accelerator and outlined regions used for the calculation of the transfer function $T(f)$. **b,** amplitude of the averaged complex transfer function T(f)/$N_i$ in units $V/Pa$ and its phase , where error bars represent maximum deviation of the respective eight independently determined trace segments highlighted in **a**.

The calibration was done using measurements performed at the Tandem accelerator. By inserting aluminum foils of 3 different thicknesses (90, 210 and 270 µm), the central energy entering the detector was gradually reduced from 9.4 MeV to 8.6, 7.2, 6.5 MeV. Since both the direct and the reflected signal contain the full information, 8 trace segments can be used for determination of the transfer function. Supplementary Fig. 2b shows the averaged transfer function whereas the error bar reflects the maximum deviation of the eight trace segments used for evaluation.  The amplitude is quite stable with small error bars. Therefor the same calibration can be used for energies in the range betweent 7 and 10 MeV. Increasing error bars beyond 10 MHz are due to the use of a 10 MHz transducer. The fact that the frequency peaks around 2 MHz while a 10 MHz transducer is employed can be explained with the effect of geometry spatial response[5]. The used transducer (Supplementary Fig. 1d) has a focal length of 25.4 mm. In our case we measured ultrasound signals with a source more than 50 mm away from the transducer. This out-of-focus operation leads to a degraded temporal resolution consistent with the measured transfer function of Supplementary Fig. 2b. An optimization of the detector response (transfer function) can further improve the resolution and I-BEAT can be adapted to the requirements.

For a quantitative calibration the measurements were also used to estimate the number of protons per proton bunch via $N_i = \int f(E_{kin}) dE_{kin} = I/\left(e f_{rep}\right)$, where $I$ is the average current that was delivered from the Tandem to the water volume, and $f_{rep} = 5$ kHz is the bunch repetition rate. We estimated, considering the pinhole size of the detector entrance and the spot size of the beam, that 60% of the bunch enters the detector such that current was estimated $= 0.6 \times 7\ nA = 5.2\ nA$ . $S_m(t)$ is thus quantitatively connected to the ideal pressure trace $p(t)$ for an arbitrary ion energy distribution $f(E_{kin})$ (predicted by equation (1)) via

$$S_m(t) = IFT[FT[p(t)]T(f)], \qquad (6)$$

## First Tests at the Tandem accelerator

Before applying I-BEAT to a laser-plasma accelerator, calibrations and first tests were performed at the MLL Tandem accelerator at Garching, using well defined proton bunches of 40 ns duration with 10 MeV ($dE/E = 10^{-4}$). For a better characterisation of I-BEAT we varied the proton energy. The initial energy of 9.4 MeV at the detector entrance was attenuated by inserting different thicknesses of aluminum in the beam path. The acoustic signals deriving from different energies were measured and are shown in Supplementary Fig. 3a, where each trace represents an average of 100 proton bunches.

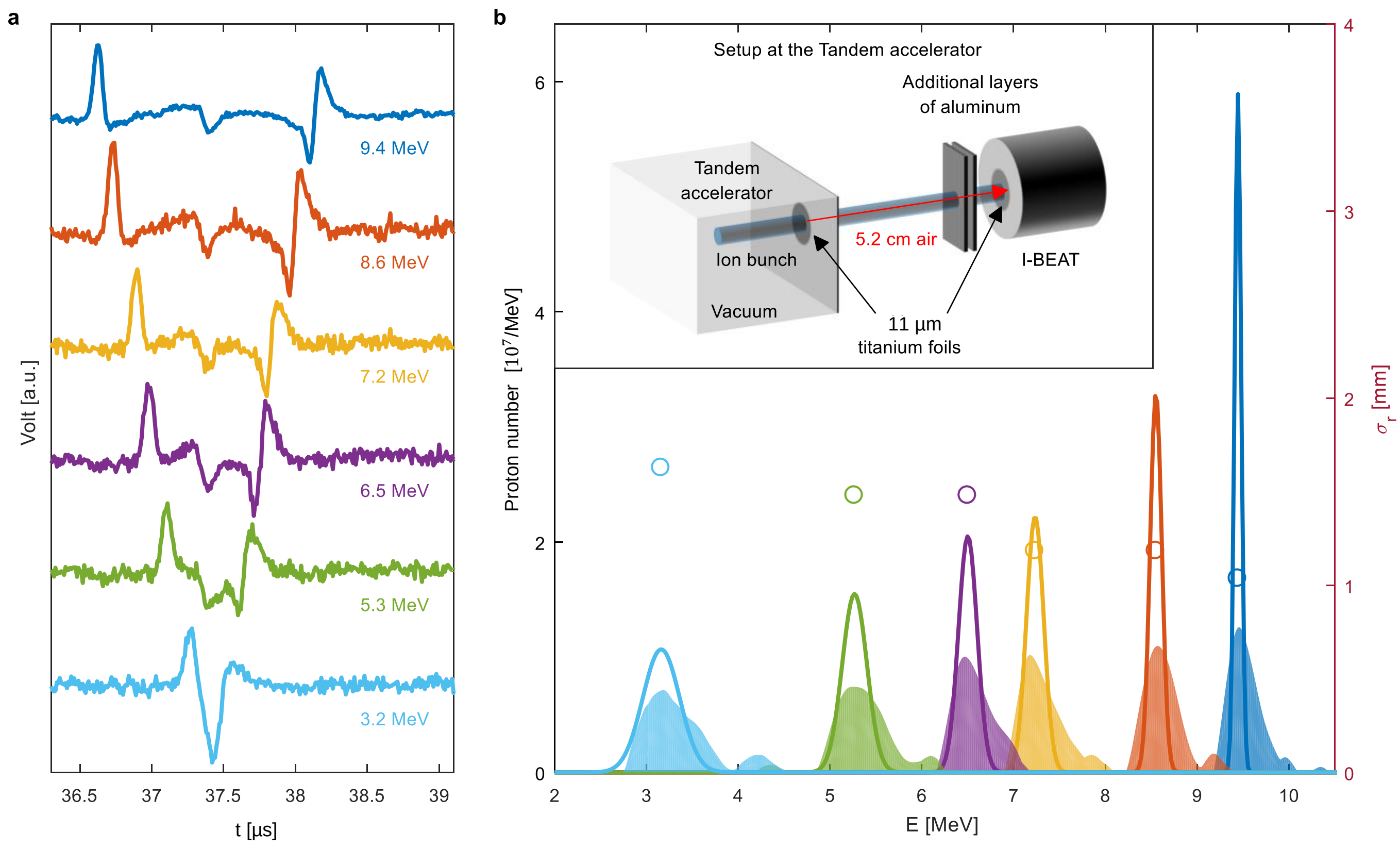

**Supplementary figure 3 | Results for Tandem-accelerated proton bunches. a**, Acoustic traces of 6 different proton bunch energies at the detector entrance recorded at the Tandem accelerator. Each trace represents an average of 100 proton bunches. **b,** Comparison of proton energy distributions simulated (curves) and retrieved data (filled), as well as reconstructed beam with standard deviation $\sigma_r$ (circles), of the measured acoustic traces in **a**. The inset of **b** sketches the setup at the Tandem accelerator.

Supplementary Fig. 3b compares the retrieved results to the energy distributions calculated via SRIM (2013)[6] by taking into account the absorber material in the beam path The theoretically expected kinetic energy distributions of protons at the position of the I-BEAT detector were calculated by SRIM, employing a mono-energetic beam with 10 MeV, passing through 11 µm titanium (the exit window of the accelerator vacuum), 5.2 cm air, and 11 µm titanium (the entrance of the water tank). While the maxima of the energy distributions (simulated and measured) agree within 1%, the larger energy spread and accordingly lower peak value of the experimental data can be explained in terms of the transfer function and its influence onto the resolution.

The transfer function expresses the limit of temporal response and thus linearly affects the longitudinal spatial resolution. Since the spatial resolution, to first order, is constant along propagation direction, the energy resolution intrinsically increases for higher ion energies and is only limited by energy loss straggling for high kinetic energies. We consider the effect of the transfer function on the FWHM of a mono-energetic ion peak to set the resolution limit. Our transfer function yields resolutions of 1.0 MeV at 5 MeV and 0.6 MeV at 10 MeV[4] (increasing resolution towards higher energies). These resolution limits of I-BEAT when applied to narrow energy spread proton bunches at low energies (< 20 MeV) are visible in Supplementary Fig. 3b. The retrieved transverse beam size $\sigma_r$ increases as expected with increasing aluminum thickness (i.e. decreasing bunch energy) due to the associated transverse straggling that becomes increasingly prominent.

**Estimation on the behavior of the detector at higher fluences**

The detector is capable of measuring really high particle fluxes. Since this scaling could not be measured so far, the expected temperature increase is estimated in this section, starting with a derivation of Boyle's law:

$$\frac{dV}{V} = -\kappa\delta p + \beta\delta T, \tag{7}$$

with kappa being the isothermal compressibility and beta the volume expansion coefficient. δp and δT are the changes in pressure and temperature respectively. As we consider only adiabatic heating in ionoacoustics, the volume expansion is neglected and only the transfer from temperature gradient to dynamic pressure is considered. With the use of the specific heat capacity, the following expression can be derived:

$$\delta p = \frac{\beta}{\kappa C_v m}\delta E, \tag{8}$$

where $C_v$ is the isochoric specific heat capacity, m the mass of the heated area and $\delta E$ the applied energy as heat. This conversion of energy to pressure is material depended and is quantified with the dimensionless Grüneisen parameter Γ:

$$\Gamma = V\left(\frac{\partial p}{\partial E}\right) = \frac{\beta}{\kappa\rho C_v} = \frac{\beta c^2}{C_p}, \tag{9}$$

where ρ is the material density, c the speed of sound and $C_p$ the isobaric specific heat capacity. In order to estimate, weather a very high particle number will significantly change the linearity of the energy transfer to dynamic pressure, the expected temperature increase and the change in the Grüneisen parameter is investigated. For liquid water and the temperature T in degrees Celsius, the Grüneisen parameter is well approximated by:

$$\Gamma_w(T) = 0.0043 + 0.0053T. \tag{10}$$

In independent measurements at the MLL Tandem accelerator, the pressure from 20 MeV protons and $3\text{x}10^6\ protons\ /mm^2$ was measured with a calibrated, broadband needle hydrophone (Precision Acoustics, UK). Measured at different distances from the source, the pressure at source level was extrapolated to 115 Pa, which in this case corresponds to a temperature gradient of 0.14 mK. Assuming a linear dependence, a temperature gradient of 4.2 K can be expected for $10^{11}$ protons/$mm^2$ per bunch.

Based on the approximation for the Grüneisen parameter given above, we derive the following values:

$$\Gamma_w(25) = 0.1368, \Gamma_w(29) = 0.158, \Delta\Gamma_w = 0.0212\ , \tag{11}$$

which corresponds to a relative change of 15%, in absolute particle numbers, over several orders of magnitude. We can thus say that the detector will also work at much higher particle numbers.

## Data Analysis

This section describes the algorithm of simulated annealing[7] in a bit more detail and discusses the influence of different beam diameters onto the retrieval in the case of the laser-accelerated ions.

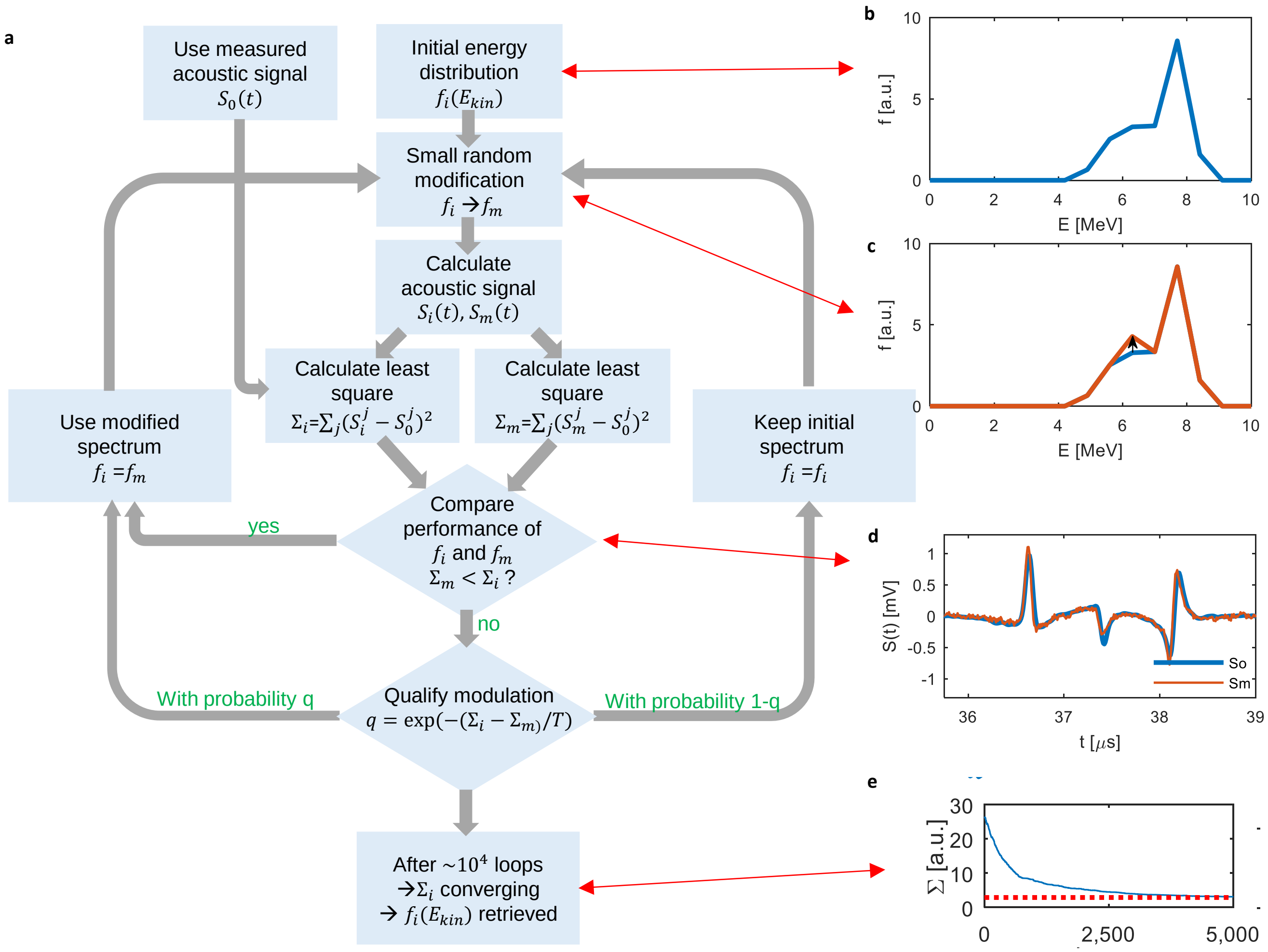

**Supplementary figure 4 | Workflow of simulated annealing**. **a,** The workflow of simulated annealing is shown. **b,** The initial spectrum $f_i$ can be guessed or started with a flat distribution. **c,** Small modulation to the initial spectrum is done ($f_m$). **d,** Acoustic signals are calculated and the performance is compared. **e,** $\Sigma_i$converges after about $10^4$ loops.

The method of simulated annealing relies on varying an initial spectrum $f_i(E_{kin})$ (a little change applies to the estimated spectrum, and both its position and amplitude are decided by pseudo-random generators) to obtain a modified spectrum $f_m(E_{kin})$. As a starting point $f_i(E_{kin})$ was chosen to be zero for all energies. Typically, the maximum of the amplitude modification is set to be smaller than 1 % of the maximum of $f_i$. With the input of the initial and the modified spectrum in eq. 1 the predicted acoustic signals $S_i(t), S_m(t)$ are calculated and compared to the acoustic signal $S_0(t)$. The residuals $\sum_i = \sum_j (S_i^j - S_o^j)^2$ and $\sum_m = \sum_j (S_m^j - S_o^j)^2$ are calculated employing the least squared method. If $\Sigma_m$ is smaller than $\Sigma_i$, the algorithm continues with the modified spectrum as the updated input distribution for the next cycle. For $\Sigma_i$ smaller than $\Sigma_m$ , with probability $q = exp(-(\Sigma_m - \Sigma_i)/T)$, the algorithm continues with the modified spectrum, while, with the probability $1 - q$, it is rejected and the initial spectrum is taken into the next cycle. This additional random choice prevents from being caught in a local minimum. $T$ is the annealing schedule temperature and was set to 1. After a sufficient amount of iterations ($\sim 10^4$) the temperature during the iteration would become stable around the temperature global minimum ($\Sigma_i$ converges), shown in the insets of Supplementary Fig. 5b, Fig. 5c and Fig. 5d , and the obtained proton energy spectrum is the retrieved spectrum. As explained before the ion bunch standard deviation $\sigma_r$ can be treated  as unknown in the retrieval process. In this case the complete process of simulated annealing depicted in Supplementary Fig. 4 is repeated by choosing another $\sigma_r$. As a result the final residuals $\Sigma_i(\sigma_r^j)$ after a sufficient number of steps (when a minimum for $\Sigma_i$) is found) shows a broad but distinct minimum for a certain bunch diameter (Fig 5a).

Fig. 2 in the main paper shows the reconstruction of laser-accelerated ion data. In this case we fixed the Gaussian bunch standard deviation $\sigma_r$ to 1.5 mm. Supplementary Fig. 5a shows the residual $\Sigma_\mathrm{i}$ in dependency of the bunch diameter for the case of the design energy set to 7 MeV. We can see that the algorithm converges for $\sigma_r$ < 2.5mm. The chosen standard deviation for the ion beam diameter of 1.5 mm can thus be explained by the look of the focus image but also the retrieval process. Fig5 b, c and d show the final result obtained for different $\sigma_r$. Also the converging residual in dependency to the number of iterations is shown as an inset. At this point the algorithm takes about 10 minutes to retrieve the complete energy spectrum of a certain $\sigma_\mathrm{r}$. Typically, the retrieval is performed for several $\sigma_\mathrm{r}$ values to find the respective optimum (depending on the knowledge of the bunch distribution). This could be improved by an advanced guess of the initial spectrum or faster algorithms.

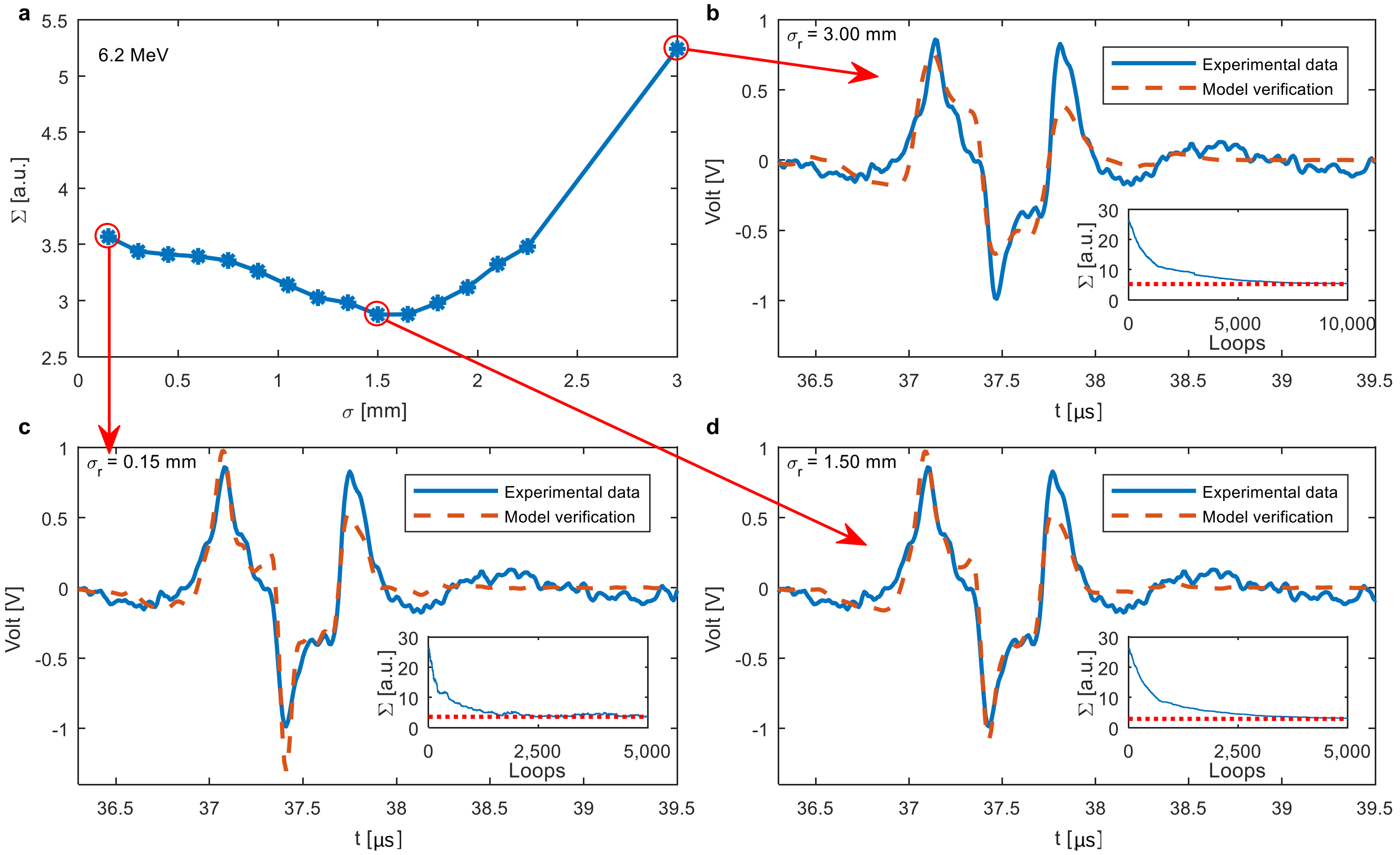

**Supplementary figure 5 | Evaluation of different ion bunch diameters for 6.2 MeV. a,** shows the residual $\Sigma_i$ in dependency of the bunch diameter. **b,c, and d,** show the experimentally recorded signal (blue) and the calculated signal using the retrieved (simulated annealing) ion energy distribution. As an inset the development of $\Sigma_i$ with the number of iterations is shown.

**The use of I-BEAT at typical conditions for laser ion acceleration experiments**

In this section we investigate the functionality of I-BEAT at different condition typically occurring at laser-ion acceleration experiments. We show the performance of I-BEAT measuring a broad band exponential spectrum that is typically obtained close to target without any manipulation of the ion bunch. We also show the performance of I-BEAT in a multi species spectrum, using quadrupoles as charge state separation.

***Functionality of I-BEAT close to target measuring a broad energy distribution***

In laser ion acceleration typically broad multispecies energy spectra emerging the plasma target[8,9]. We simulated the performance of I-BEAT positioned close to target without any manipulation of the energy distribution (e.g. magnetic quadrupoles). The proton input spectrum was exemplarily taken from[10]. Note that other ion species are typically emitted with significant lower particle numbers and energy and are thus neglected in this consideration. Without the use of charge state separating fields a differentiation of different ion species is not possible. Assuming and opening aperture for the detector with an radius of 3 mm (seems feasible since it supported by the measured data) covers an area of about A = 30 $mm^2$ and thus the I-BEAT detector was positioned such that $10^9$ protons reach the detector. The particle number was chosen to obtain a good signal to noise ratio (knowing the pressure signal of a single proton and the background noise). The measured spectrum provides more than $10^8$ protons per msr (all energies summed up). We thus have to cover a steradian Ω of 10 msr. With $\Omega = \frac{A}{d^2}$ and *d* being the distance to the detector yields d = 50 mm and has thus be positioned close to the target. The given input spectrum was used for a calculation of the expected signal (Supplementary Fig. 6 b). This signal was then evaluated with the I-BEAT algorithm and thus the spectrum was evaluated. In Supplementary Fig. 6a the original spectrum is compared to the ones evaluated with I-BEAT. The required signal to noise ratio thus sets a limit to the distance (to the source), where the I-BEAT detector can be placed. I-BEAT can thus function as the typical RCF stack that is positioned close to target, admittedly without measuring the beam profile but offering an online evaluation.

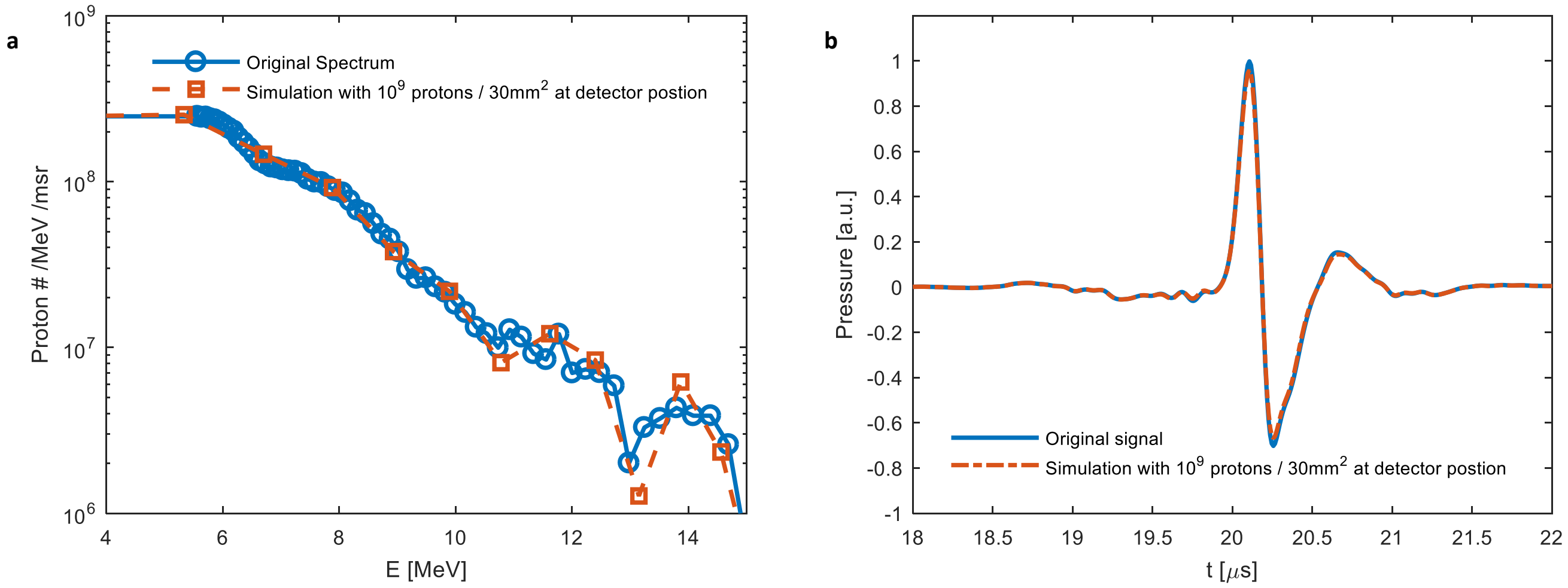

**Supplementary figure 6 | I-BEAT positioned close to target. a,** Exemplary measured TNSA spectrum taken from[10] (blue curve) and the reconstructed spectrum assuming $10^9$ protons at the detector (red curve). **b,** Calculated pressure signal. The blue curve is without any noise disturbance while the red curve is evaluated for $10^9$ protons at the detector incorporating the measured noise level. The signal to noise ratio at this high particle number is good and allows the reconstruction of the broad band energy distribution.

**Functionality of I-BEAT measuring multiple species**

Laser-driven plasmas accelerate not only protons but also other ion species depending on the target material. Especially carbons at different charge states are also emitted from the contamination layer of any target surface. Since the range in the water tank is dependent on the mass (not so much on the initial charge state) and the kinetic energy, I-BEAT is able to assign a certain energy to a certain charge and mass in combination with an energy selective focusing device[11,12] (such as magnetic quadrupoles) and can thus, at least in this configuration also reconstruct the energy distribution of different ion masses and charges in a single shot. An example calculation is presented in the Supplementary Fig. 7. We assume flat spectra of carbon ions with charge states 4, 5 and 6, as well as protons. The QP-doublet focusses all ions with the same synchrotron radius to the same point as depicted in Supplementary Fig. 7a. A calculation of the signal when such a multispecies ion bunch is measured with I-BEAT is performed and the expected acoustic signal is shown in Supplementary Fig. 7b and 7c. One can clearly distinguish the contributions of the different ions to the acoustic wave form and hence measurement of this waveform will allow for reconstructing the complete information. Of course, the information of the complete ion spectra emitted from the target remain inaccessible (as the QPs filter out ions which are too far of the design energy which is focused). I-BEAT will not replace the currently and also really valuable techniques of characterising the composition of the ion-spray emitted from the target, such as provided by Thomson parabola spectrometers[13,14]. But it will be an additional and complementary option to measure ion energies that will give an experimentalist a new, very powerful, tool for future research at the application site of high flux ion bunches.

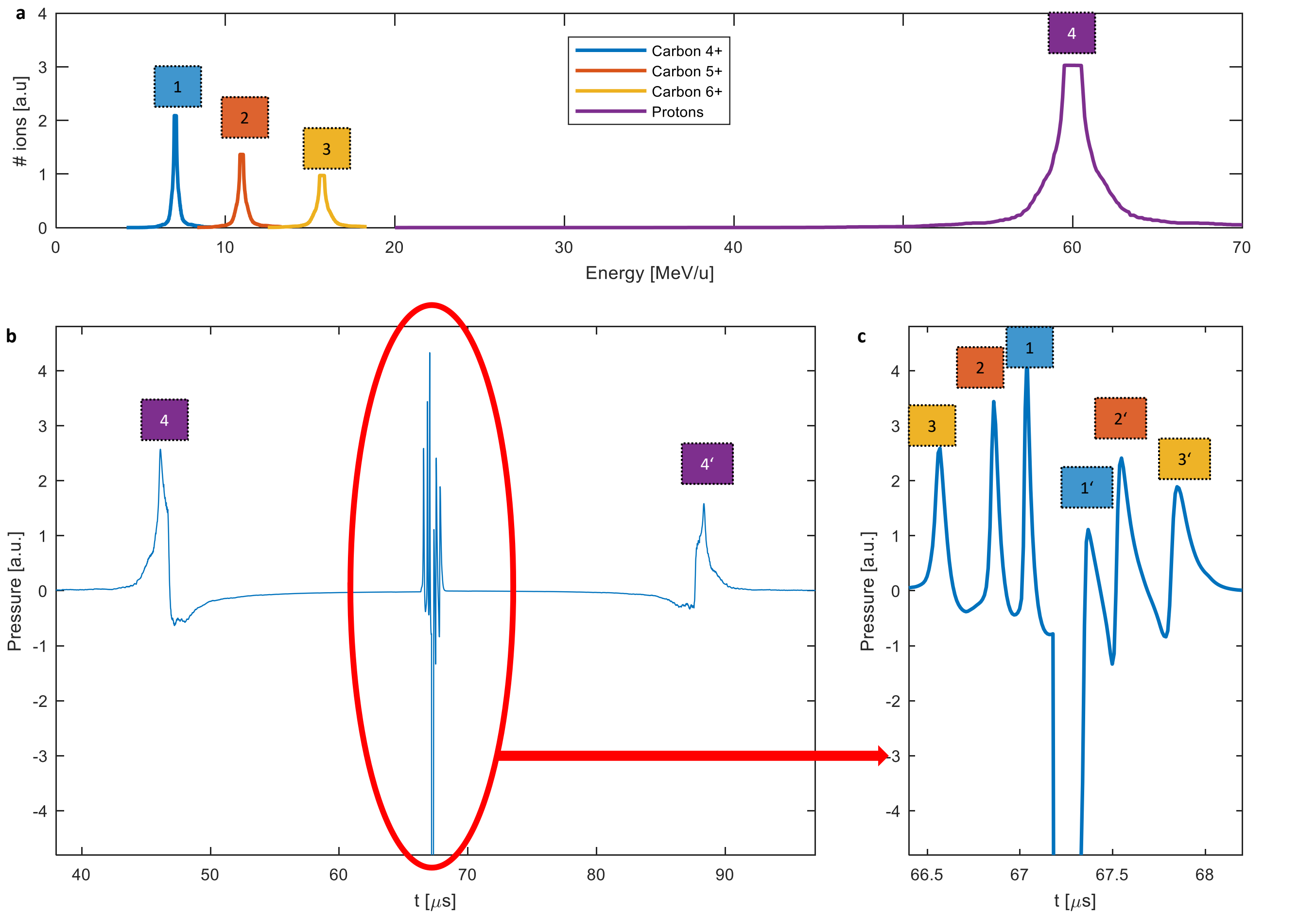

**Supplementary figure 7| Multispecies in combination with magnetic quadrupoles. a,** Multispecies ion energy distribution selected by quadrupoles, set to a design energy of 60 MeV protons. **b,** Simulated acoustic trace generated by the spectrum in **a**. **c,** Enlargement of the central part of **b** (highlighted with red). The carbon ions do not penetrate far into the water but are still well separated in the oscilloscope trace. Since the peaks are well separated, I-BEAT is capable of reconstructing the complete information.

# References for the Supplementary Material